# Magnetic Resonance Imaging of Single Organic Radicals with Sub-Molecular Resolution

Gregory Czap[1], Christoph Wolf[2,3], Jose Reina-Gálvez[2,3,4], Mark H. Sherwood[1] and Christopher P. Lutz[1*]

[1]IBM Almaden Research Center, 650 Harry Road, San Jose, CA 95120, USA

[2]Center for Quantum Nanoscience (QNS), Institute of Basic Science (IBS), 03760 Seoul, Republic of Korea

[3]Ewha Womans University, 03760 Seoul, Republic of Korea

[4]Department of Physics, University of Konstanz, D-78457 Konstanz, Germany

*Corresponding author. Email: cplutz@us.ibm.com

## Abstract

Interest in the magnetism of organic compounds is growing because of new organic magnets, spin-based electronics and the diverse properties of magnetic edge states in graphene nanoribbons. Electron spin resonance spectroscopy combined with the scanning tunneling microscopy has recently been developed as a powerful tool to address individual magnetic atoms and molecules at the atomic scale. Here we demonstrate electron spin resonance and magnetic resonance imaging of all-organic radical anions adsorbed on a protective thin insulating film grown on a metal support. We show that using the highly localized exchange field of the magnetic tip apex allows visualization of the delocalized spin density with sub-molecular resolution, enabling spin-density tomography that can distinguish similar molecular species. These results provide new opportunities for visualizing spin density and magnetic interactions at the atomic scale.

## Introduction

The magnetism of organic radicals has long been used in bulk electron spin resonance (ESR) of reaction intermediates and of spin-labeled molecules that serve as a probes of molecular structure and conformation (*1*). Interest in organic molecular magnetism has intensified recently for spin-based molecular electronics (*2*), magnetically ordered arrays (*3*, *4*), molecular qubits (*5*), and magnetism in nanographene radicals (*6*, *7*), which show topological frustration and intramolecular spin coupling (*8*, *9*). Their spin properties have been sensed and imaged with sub-molecular resolution by spin-polarized tunneling (*7*), Kondo resonance detection (*10*, *11*), and inelastic electron tunneling spectroscopy (IETS) (*10*), but thermal broadening limits the energy resolution of these methods.

The study of magnetism at the atomic scale has significantly benefitted from the combination of ESR with scanning tunneling microscopy (ESR-STM), which together offer very high energy and spatial resolution, and coherent control of individual quantum spins (*12*, *13*). In this technique, a radio frequency (RF) voltage $V_{RF}$ is applied to the tunneling junction (*12*) to drive spin resonance. The change in the tunnel current on-resonance is then used to sense the spin state by magnetoresistance. The ~20 neV spectral resolution of ESR-STM combined with the capability to coherently manipulate spins and to position atoms and molecules provide new opportunities to study nanomagnetism, quantum sensing and quantum information in atomic-scale structures (*14*). Spin centers on transition metals (*15*), organometallic compounds (*16*), alkali dimers (*17*), and lanthanide atoms (*18*, *19*) have been studied in ESR-STM. The delocalized unpaired electrons in π orbitals of organic molecules offer a vast set of molecular spin candidates (*6*, *20*), and have recently been detected by ESR in STM (*21–23*) and in AFM (*24*).

Here we study organic molecular magnetism by detecting ESR and the spatial distribution of the unpaired spins in organic radical anions. These radicals were adsorbed on a thin film of epitaxially grown MgO in order to provide controlled isolation from the conduction electrons of the supporting Ag(001) substrate (*25*). We study several derivatives of fluorene, a small representative polycyclic aromatic hydrocarbon (PAH) whose radical derivatives have been studied in ensemble ESR (*26–28*) and scanning probe microscopy (*29*, *30*), and employed for molecular wires (*31*) and magnetic chains (*3*). We find that these molecular adsorbates have high electron affinity, spontaneously charging to become anions on this surface. This charging quenches neutral radicals, and creates radical anions from non-magnetic neutral species (*32*). ESR-STM measurements on the radical anions demonstrate that their delocalized molecular spins can be driven coherently. We present scanning magnetic resonance imaging (MRI) (*17*, *33*) of these molecules, and show that the highly localized magnetic exchange field from the tip allows direct three-dimensional visualization of the regions of spin accumulation within the radical. The ultra-high spatial resolution combined with neV energy resolution allows atomic height corrugations to be mapped with the MRI images so that similar molecular species are readily distinguished.

## Results

The molecules were sublimed onto a cold (~7–10 K) film consisting of two atomic monolayers (2 ML) of MgO grown epitaxially on an Ag(001) substrate (see Methods and SM for details). Fe and Ti adatoms were co-dosed by evaporation onto the cold surface. The Fe facilitates spin-polarized Fe tip preparation, and the Ti provides a well-studied spin $S = \frac{1}{2}$ reference adsorbate. All data shown was acquired at temperature $T = 1.1$ K.

Molecules of 4,5-diaza-9-fluorenone (DAF) (*34*) (**1**), 9-fluorenone (fluorenone) (**2**), and 2,7-dibromo-9-fluorenone (DBF) (**3**) were co-adsorbed onto the MgO film (Fig. 1) to provide a comparison of structurally similar molecules. All three are closed-shell and therefore non-magnetic when in the neutral state, but high electron affinity that results a radical anion when adsorbed on this surface. A related molecule, 9-bromo-fluorene (**4**), was deposited with the intention to explore the debrominated radical, 9-fluorenyl, as a representative dangling-bond species. The 9-bromo-fluorene adsorbed both in the bromine "down" orientation, with the bromine atom oriented toward the MgO surface (**4**a), and bromine "up", oriented away from the surface (**4**b). These molecules were dissociated by voltage pulses applied to the junction with the tip positioned over each molecule, causing them to dissociate into 9-fluorenyl (**5**) and a bromine atom that was found to have moved typically 1–2 nm away on the surface.

The compounds were dosed one at a time in order to confidently identify each molecule and its orientation in the plane of the surface by its appearance in STM images (Fig. 1B,C). In addition to conventional STM images, structural images using a CO-terminated tip were acquired of fluorenyl and DAF adsorbed on the bare metal Ag(001) surface by using the inelastic tunneling probe (itProbe) technique (35, 36) (Fig. 1E,F). In this method, the soft lateral hindered translation vibration mode of the CO on the tip is used to sense short-range repulsive forces in a manner similar to CO-tip non-contact atomic force microscopy imaging of chemical structures (37–40). The itProbe images (Fig. 1E,F) confirm the chemical identity of fluorenyl and DAF. The ketone oxygen of DAF appears as a repulsive ridge oriented orthogonal to the C–O bond axis (Fig. 1F).

We investigated whether the adsorbed molecules carried unpaired spins by using IETS to detect spin-flip excitations (see SM), and by ESR detection of electron spins. The DAF, fluorenone, and DBF species all showed spin $S = ½$ magnetic behavior in both IETS and in ESR. In contrast, the fluorenyl and atomic bromine dissociation products were found to be non-magnetic. These observations and density functional theory (DFT) calculations indicate that each species spontaneously charges to an anion upon adsorption. This charging behavior for small adsorbates on metal-supported insulating films has been observed before for high electron affinity species charging to anions (16, 23), as well as low ionization energy species charging to cations (17, 32, 41). Accordingly, the spins of radical species fluorenyl and atomic bromine are quenched upon charging from adsorption on this surface, while the neutral closed-shell molecules (DAF, fluorenone, and DBF) become radicals upon charging.

ESR spectra acquired on DAF with the tip positioned at three representative lateral positions above the molecule show resonant peaks corresponding to an electron spin having a $g$-factor of ~2 (Fig. 2). The peaks shift with the height of the tip, which was varied at each position by adjusting the setpoint tunnel current $I_{set}$. The peaks shift because of the effective magnetic field of the tip $B_{tip}$ at the position of the molecule, which arises from a combination of exchange interactions and dipole coupling (42). Extrapolating the resonant frequency to zero current gives the adsorbed molecular spin's intrinsic (tip-independent) Larmor frequency, which yields a $g$-factor of $1.97 \pm 0.04$ for DAF and $1.98 \pm 0.03$ for fluorenone. At each lateral tip position, the resonance frequency decreases as the tip approaches the molecule, demonstrating that for this tip, $B_{tip}$ partially cancels the applied magnetic field $B$. This behavior is consistent with a tip spin aligned to $B$, together with antiferromagnetic coupling to the molecule. Frequency shifts greater than ~1 GHz can be attributed primarily to interaction with the exchange field of the tip (43), and the sign of this exchange is the same at each of the three locations shown in Fig. 2. Some tips (such as the one used in Fig. 2) gave shifts to lower frequency as the tip was lowered, while some shifted the resonance to higher frequency (Fig. 3I–K). For each tip, however, the sign of the shift was the same on all parts of the molecule, regardless of lateral position.

To investigate the dependence of the molecule's ESR signal on the lateral position of the tip, we acquired MRI images by measuring the ESR signal at a fixed RF frequency $f$, while scanning the tip in a plane parallel to the surface (Fig. 3). The RF frequency was detuned from the molecular spin's intrinsic Larmor frequency, which is determined by the molecule's $g$-factor and the externally applied $B$. The resulting images show narrow ring-shaped regions, where the transition frequency of the unpaired spin was shifted precisely into resonance under the influence of magnetic coupling with the tip. These "resonant slices" are conceptually identical to those observed in magnetic resonance force microscopy measurements of single spins (*44*). Here, the close proximity of the tip results in strong interactions and atomic-scale spatial resolution. Previous imaging of tip-molecule exchange interaction using IETS achieved sub-nanometer resolution, but thermal broadening limited the energy resolution and consequently the spatial resolution (*45*, *46*).

Each MRI image measures a two-dimensional plane that samples the three-dimensional magnetic interaction isosurface, resulting in roughly circular rings where the plane and isosurface intersect. These resonant slices reveal sub-molecular features that can be understood by considering the molecule's spin density (Fig. 3G), which is determined largely by the unpaired spin occupying the singly-occupied molecular orbital (SOMO) of the radical anion (Fig. 3F). The SOMO is delocalized over the entire molecule but is concentrated in orbital lobes. The resonant slices thus form rings of constant interaction energy that enclose lobes of high spin density, demonstrating that the tip field has high enough spatial resolution to selectively interact with individual lobes.

To investigate the resonant slice patterns, we acquired sequences of MRI images using two different schemes: by varying the applied RF frequency $f$ in successive images (Scheme 1, Fig. 3A), and by varying the tip-surface distance $z$ in successive images (Scheme 2, Fig. 3B). Varying $f$ varies the size of the interaction energy isosurfaces, resulting in larger rings as the frequency is increased to approach the intrinsic resonant frequency $f_0$ (Fig. 3C–E). These images correspond to mapping different magnitudes of $B_{tip}$ sensed by the molecule as the tip is laterally scanned, where $f$ closer to $f_0$ corresponds to smaller $B_{tip}$, which in turn yields larger rings because resonance occurs when the tip is laterally farther away from each spin lobe. The images confirm that the rings result from ESR because only $f$ is changed between images.

When varying the tip-surface distance (Scheme 2) while leaving $f$ unchanged, different MRI images correspond to probing a fixed tip-molecule interaction isosurface at different heights (Fig. 3I–K,O–Q). For these MRI sequences we chose an RF frequency such that the tip field at which spin resonance takes place is held at a relatively high value (~90 mT), where exchange interactions are expected to dominate. With a suitably chosen tip apex, exchange interactions are nearly isotropic and thus are expected to produce nearly circular resonant slice rings as observed. For both DAF and fluorenone, the spatial patterns of the resonant slice rings show strong similarity to the DFT-calculated spin density of the free (gas-phase) molecular anions (Fig. 3G,M). The resonant slice rings surround regions of high spin density, most prominently at sites A, B, and C as labeled in Fig. 3G,M. This MRI method thus provides a means for profiling the spin density with submolecular resolution.

The MRI images for DAF (Fig. 3I–K) are qualitatively similar to those for fluorenone (Fig. 3Q–O), but a prominent difference is the order in which resonant slices appear at each spin lobe. As the tip was lowered from large tip-surface separation, a resonant slice ring appears in DAF first at site A (Fig. 3I). As the tip height reduced further, the site-A ring increases in radius, and additional rings appear near sites B and C (Fig. 3I–K). In contrast, a ring appears first at site C in fluorenone (Fig. 3O), with additional rings appearing near sites A and B as the tip is lowered. This behavior was reproduced with most other tip apexes, despite tip-dependent variation in detailed properties.

The DFT-calculated spin density distribution of the gas-phase DAF anion is nearly identical to that of the gas-phase fluorenone anion (Fig. 3G,M). They are structurally similar but their adsorption geometries on the MgO surface, such as tilting and bending, may differ because of the electronegativity of the N atoms in DAF. These geometric differences result in differences in the MRI images as we now discuss.

We performed DFT calculations for both fluorenone and DAF adsorbed on the Ag-supported MgO film. The lowest-energy relaxed geometry for both was found to be in the "diagonal" orientation relative to the MgO oxygen rows (Fig. 4A,E) which is also the most frequently observed orientation in the experiment. Upon adsorption, DFT shows that each molecule naturally forms an anionic radical by charge transfer from the Ag substrate. In their relaxed positions, the molecules were found to tilt with respect to the surface, and to distort from their gas-phase planar geometry (Fig 4B,F). For both DAF and fluorenone, the ketone oxygen moves toward the surface relative to the rest of the carbon skeleton, as for PAH ketones on other surfaces (47). The corresponding spin density on the oxygen atom similarly moves toward the surface and away from the tip, likely accounting for the absence of spin-resonant contrast near site E (the oxygen atom) in MRI images (Fig. 3). The two species show significant differences in how the carbon skeleton tilts and deforms. For fluorenone, the molecule remains largely planar but with a tilt angle of ~6° with respect to the MgO surface plane. This causes the spin density near site C (the spin lobe farthest from the ketone) to protrude away from the surface. For DAF, this tilting of the carbon frame is suppressed by the attractive interaction between each nitrogen atom and the Mg atoms of the MgO lattice. This attraction is evidenced also by a ~14 pm lifting of two Mg atoms from the surface toward the N atoms of DAF, which is completely absent in the case of fluorenone. Consequently, the spin density near site A (ketone carbon) protrudes away from the surface the most for DAF (Fig. 4), whereas for fluorenone the overall tilt of the carbon skeleton causes the site C spin density to protrude more than site A. In summary, the DFT calculated adsorption geometries of the two molecular species provide an explanation for the MRI image features and show that MRI tomography reveals information about three-dimensional molecular geometry with submolecular resolution.

We expect that the exchange interaction between the tip and molecule is nearly isotropic and with a magnitude proportional to the overlap between the tip spin and each spin lobe of the delocalized radical. This interpretation accounts for the prominent appearance of resonant slices in MRI images associated with spin centers at sites A, B, and C, but resonant signals near the high spin density region at site D were only rarely observed (Figs. 3Q and 4G), and none were seen at site E. The adsorption geometry explains the lack of an MRI signal at site E, near the ketone oxygen, because it is bent toward the surface and away from the tip. However, the geometry does not account for the absence of MRI features near site D. To explain this absence, we note two factors. First, the spin density is negative (minority spin orientation) on the nearby N atoms in DAF, and in the corresponding carbon atoms in fluorenone (Fig. 3G,M and Fig. 4A,E). This small negative spin region is expected to reduce the total exchange integral between the tip and molecular spins when the tip is positioned near site D. Second, the exchange integral depends on the wavefunction phase of the delocalized molecular spin. Since the SOMO lobe at site D is surrounded by other SOMO lobes with opposite quantum phase (sites A, C) that have some overlap with the tip apex, the total exchange integral is reduced. Despite the near absence of resonant slice rings on DAF and fluorenone near site D, the corresponding site-D resonance was observed clearly in related molecules: DBF, and fluorenone rotated in the surface plane to the "horizontal" orientation.

A striking feature of the MRI images is the reversal of the sign of the ESR signal on different parts of the molecule, showing light or dark rings at different regions. For most tips tested, the resonant slices in surrounding sites A and C of both DAF and fluorenone were light (higher conductance when on-resonance) while region B slices were dark (lower conductance). This reversal of contrast agrees with the point spectroscopy results (Fig. 2), where ESR peaks occur at sites A and C, and dips occur at site B. We also observed cases of contrast reversal in a single resonant slice ring, as in Fig. 5H. In previous studies of adsorbed transition metal atoms (*15*, *33*), ESR spectra acquired of Ti usually showed an increase in current (peak) on-resonance whereas spectra acquired on Fe atoms showed a decrease in current (dip). This behavior was attributed to the sign of the tunneling magnetoresistance coefficient of each atom (*15*) and was assumed to be an intrinsic property of the adsorbate, but no mechanism was discussed that could account for the lateral spatial variation in the tunnel magnetoresistance observed here.

We propose that the inversion of MRI contrast between lateral tip positions originates from a change in the dominant tunneling channel, which occurs primarily through the SOMO orbital for some lateral regions of the molecule and the SUMO for other regions, resulting in reversal of the magnetoresistive contrast. At the bias voltages used here tunneling occurs in the Coulomb blockade regime so that co-tunneling processes dominate, which can primarily have either SOMO-like or SUMO-like character. Majority-spin electrons tunnel readily through the SOMO, giving a large tunnel current when the molecule's spin is aligned with the tunneling electrons from the tip. This results in a dip in the ESR spectrum because spin resonance partially flips the spin, reducing the magnetoconductance. Similarly, only minority-spin electrons tunnel readily through the SUMO (Fig. 5A,B), resulting in an ESR peak. We note that the change in sign of the magnetoresistive ESR signal is not expected to occur when the SOMO and SUMO are equidistant in energy from the Fermi level, but instead at an offset value due to interference between spin-polarized and non-spin-polarized transport channels (*48*, *49*) (see SM).

We used a non-equilibrium transport model including DC bias and RF voltages to develop understanding of the ESR spectra. In this model, a single orbital Anderson impurity represents the molecular spin, which is connected to a spin-polarized electrode (the tip) and a non-polarized substrate electrode. When the hopping operator of the polarized electrode is modulated on-resonance with the Zeeman-split molecular spin states, this model predicts an ESR signal in the long-time limit measurable in the DC current detected in ESR-STM (*50*). As shown in Fig. 5E, F, the simulation reproduces most of the main features seen in the measurements, including the trends in both peak amplitude and frequency shift as a function of the bias voltage. This frequency shift has been attributed before to piezoelectric motion-driven modulation of the adsorbate *g*-factor (*51*), whereas in the present work, our simulations show the bias voltage modulates the virtual hopping rate and accordingly the tip's kinetic exchange field (*52*). Although the transport model produces overall excellent agreement with experiment, a notable exception is the inversion of contrast at Site B. We attribute this reversal to co-tunneling through the SOMO instead of the SUMO as is the case at Sites A and C, and the associated reversal of magnetoresistance is not captured in the sequential tunneling model used here (see SM for a detailed discussion).

There are several mechanisms that could lead to dominance of either the SOMO or the SUMO tunneling channel at a given lateral tip position: the energy of each orbital relative to the Fermi level, the bias voltage, local electrostatic gating by the STM tip, and the intrinsic spatial wavefunction of each orbital. We discuss each of these in turn. The energies of each orbital could not be measured directly because they fell outside the $|V_{DC}| < 0.5$ V energy range accessible in experiment, and larger voltages caused the molecules to hop laterally or to change orientation. DFT calculations indicate that the SOMO and SUMO lie ~1–1.5 eV from the substrate Fermi level, which results in roughly equal contributions to tunneling at low bias voltages. The STM tip is known to provide localized electrostatic gating, which shifts the orbital levels relative to the Fermi level, potentially shifting the dominant transport channel towards either the SOMO or SUMO energy levels (Fig. 5A,B). The tip gating effect varies as the tip is repositioned laterally over adsorbates, and has been shown to be large enough to alter the charge state of single molecules with sub-molecular spatial sensitivity (*53–55*). Consequently, shifting of the SOMO and SUMO levels relative to the Fermi level can result in a reversal of the sign of magnetoresistance so long as the dominant tunneling pathway switches between SOMO and SUMO cotunneling. We propose that tip gating shifts the SOMO and SUMO levels downward with respect to the Fermi level such that the SUMO channel is preferred when tip gating is strong (*55*). This interpretation implies that near zero bias voltage the electric field is directed from the tip to the sample. DFT calculations performed for a model Fe tip consisting of a pyramidal arrangement of 5 Fe atoms on a Ag(001) slab indicate that there is indeed a positive electric field from the tip across the vacuum gap (see SM). This agrees with previous calculations for model Cu tips, where the positive charge accumulation at the tip apex was attributed to the Smoluchowski effect (*56*). In a similar manner, the bias voltage also acts in concert with the tip gating to favor either SOMO or SUMO tunneling (Fig. 5A–F) as seen in the strengthening of SUMO-detected ESR rings (sites A and C) when going from negative to positive bias, and a corresponding weakening of the SOMO rings at site B (Fig. 5H,G).

In addition to the above factors, SOMO and SUMO resonances may differ in their spatial wavefunctions, as demonstrated for singly charged adsorbates (*57*). This difference arises from the different Coulomb potential that the electron in the SOMO experiences compared to the second electron added when tunneling through the SUMO, which sees an additional unscreened electron that modifies the spatial wavefunction. We propose that the individual lobes in the SOMO and SUMO wavefunctions vary in magnitude with respect to one another, which influences the site-dependent cotunneling rate through each channel, resulting in different magnetoresistive contrast as the tip is repositioned at each lobe on the molecule. A more detailed discussion of the site-dependent ESR signal reversal, including other plausible mechanisms such as spin-torque driving (*23, 58*) is presented in the SM.

We observed that the MRI contrast reverses on all parts of the molecule at negative bias voltage relative to positive bias (Fig. 5H), where we consider the change in absolute current on resonance defined as $\Delta|I| = |I_{RF\ ON}| - |I_{RF\ OFF}|$, such that lighter regions indicate higher conductance. This additional reversal mechanism, which is a separate phenomenon from the tip-position-dependent reversal, was observed for all tunnel conditions tested, and for all lateral positions of the tip above the molecule. At negative bias, spin torque initialization by the spin-polarized tunneling electrons incoherently initializes the molecule to an off-resonance spin orientation opposite the Zeeman ground state, which then reverses the magnetoresistive contrast on resonance (*23, 59*) (see SM).

The results shown here demonstrate a method to visualize the orbitals of delocalized spins in organic radicals with atomic-scale resolution. The MRI scans of organic radicals reveal resonant slices that map the spin density of their delocalized π electrons in real space with unprecedented spatial resolution. The high spectral resolution of ESR-STM results in correspondingly high spatial resolution in all three dimensions, allowing tomographic visualization of spin density. This three-dimensional characterization discerns differences in the atomic height corrugations within closely related molecules. The MRI scans also provide a direct visualization of where to position the STM tip to most effectively drive and sense the delocalized spin. These results should be broadly applicable to the use of ESR-STM in the study of organic magnetism, including magnetism in widely-studied systems such as graphene nanoribbons, open-shell nanographenes, and planar polycyclic aromatic compounds. This spin-exchange mapping may allow high-energy-resolution atomic-scale imaging of complex magnetic ordering and interactions such as anisotropic exchange, as well as dipolar and intra-molecular coupling.

## Methods

### STM and sample

Experiments were performed in a home-built ultra-high vacuum STM operating at $T = 1.1$ K equipped with a superconducting magnet oriented nearly along the surface plane (~8° out of the plane). This orientation was chosen for previous experiments. RF was applied to the tip through semi-rigid coax cable to the STM, with constant-amplitude RF frequency sweeps obtained as in previous work (*60*).

The Ag(001) single crystal was prepared by cycles of sputtering in $4 \times 10^{-6}$ Torr Ar followed by annealing to 500 K. The 2-ML MgO thin film was grown epitaxially on the clean Ag(001) substrate at the rate of ~0.5 ML/min by holding it at ~340 K while evaporating Mg onto the surface from a Knudsen cell in a $1 \times 10^{-6}$ Torr $O_2$ environment. The MgO-covered sample was then transferred to the cold STM.

The bulk Ir tip was presumably Ag covered near the apex, and was prepared by 10 V field emission followed by plunging ~0.2–1 nm into bare Ag regions. Spin polarized tips were then prepared by transferring ~2–8 Fe atoms to the apex from the MgO. Tips are considered distinct when they differ in their atomic arrangement at the apex.

Br-terminated tips were prepared by transfer of a Br atom to the metal apex by positioning the tip above an MgO-adsorbed Br or a 9-bromo-fluorene molecule with Br "up" and lowering the tip ~0.2 nm from a setpoint of 100 pA at 100 mV. Br tip terminations were found to sharpen the STM imaging contrast, reduce the conductance of the tip, and reduce the probability of unintentionally transferring a molecule to the tip at close tip-surface distance. Fe-Br tips were often spin polarized and facilitated ESR measurements despite the Br tip termination, but we observed it was more common to encounter magnetically bistable tips when prepared this way. The Br was removed from the tip by positioning the tip above a metal atom such as Fe or Ti at and lowering it ~0.3–0.35 nm from a setpoint of 100 pA at +100 mV, resulting in a metal-Br complex on the surface. We often observed that functionalizing a metal tip with Br and then removing the Br through deposition onto a metal adatom enabled recovery of the previous metal tip apex without changes to its structure.

### Deposition

The organic molecules were sublimed onto the clean 2-ML MgO/Ag(001) substrate held at $T$

= ~7–10 K. The molecules were purified by sublimation and admitted to the room temperature vacuum chamber through a leak valve, monitored with a mass spectrometer, and a shutter was opened to deposit them onto the surface at the cold STM.

Fe and Ti were dosed by e-beam heating pure metal targets in the room-temperature, with a shutter opened for line-of-sight to the cold sample surface. Some publications indicate that the Ti atoms may be hydrogenated (forming TiH)(*43*, *15*, *61*) and we describe them here as Ti for simplicity.

**DFT calculations**

Freestanding (gas phase) DFT calculations were performed using GAMESS (*62*) at the B3LYP/6-31++G level.

Plane wave DFT calculations were performed using Quantum Espresso 7.1 (*63*, *64*). We used pseudopotentials from the PSL library 1.0.0 (*65*) and a cutoff of 60 Ry for the kinetic energy and 600 Ry for the charge density. All calculations use the gamma point only but we checked for convergence using 2×2×1 and 3×3×1 *k*-grids for the slab calculations and found no significant difference. Dispersive forces were treated using the revised VV10 (rVV10) functional (*66*). For calculations of the isolated molecules the molecules were placed in a 2×2×2 nm vacuum box and relaxed until the total force was less than 0.01 Ry/$a_0$ and energy differences were less than $10^{-3}$ Ry. For calculations of the adsorbed molecules, we built lateral supercells (*a* = 1.664 nm) of Ag (100) capped by 2 ML of MgO and more than 1 nm of vacuum. The gas phase molecule was then placed at ~0.4 nm height and relaxed to obtain the ground state. We compared several rotational orientations of the molecules to ensure we found the ground state using nudged elastic band calculations.

**Transport model for ESR spectra**

Transport simulations were performed using non-equilibrium transport including the RF contribution in the barrier modulation model (*50*). In this model, the adsorbate is an Anderson impurity with ionization energy $\varepsilon = -1.22$ eV and Hubbard $U = -2\varepsilon = 2.44$ eV, in agreement with the absence of detected electronic resonances in the ±400 mV range that was accessible for tunneling spectra (d$I$/d$V$) of these adsorbates. SOMO (SUMO) transport was modeled by gating of the impurity with respect to the electrode levels by $-1$ (+1) eV relative to the Fermi level. Temperature was set to $T = 0.5$ K, tip polarization was $P = 0.5$, and driving of 25% with respect to the coupling to the tip which was set to 14 μeV, while the coupling to the substrate was three times larger. The gating value is in line with DFT simulation and with this value we fit $\varepsilon$, $U$ and the coupling to the tip from ESR resonance position using the exchange field theory in references (*67*, *52*, *68*).

## Supplementary Materials

Supplementary Materials (SM) are not available for this version of this manuscript.

# References

1. J. A. Weil, *Electron Paramagnetic Resonance: Elementary Theory and Practical Applications* (Wiley-Interscience, Hoboken, N.J, 2nd ed., 2007).

2. E. Coronado, Molecular magnetism: from chemical design to spin control in molecules, materials and devices. *Nat Rev Mater* **5**, 87–104 (2019).

3. X. Lu, S. Lee, J. O. Kim, T. Y. Gopalakrishna, H. Phan, T. S. Herng, Z. Lim, Z. Zeng, J. Ding, D. Kim, J. Wu, Stable 3,6-Linked Fluorenyl Radical Oligomers with Intramolecular Antiferromagnetic Coupling and Polyradical Characters. *J. Am. Chem. Soc.* **138**, 13048–13058 (2016).

4. K. Sun, N. Cao, O. J. Silveira, A. O. Fumega, F. Hanindita, S. Ito, J. L. Lado, P. Liljeroth, A. S. Foster, S. Kawai, On-surface synthesis of Heisenberg spin-1/2 antiferromagnetic molecular chains. *Sci. Adv.* **11**, eads1641 (2025).

5. F. Lombardi, A. Lodi, J. Ma, J. Liu, M. Slota, A. Narita, W. K. Myers, K. Müllen, X. Feng, L. Bogani, Quantum units from the topological engineering of molecular graphenoids. *Science* **366**, 1107–1110 (2019).

6. D. G. De Oteyza, T. Frederiksen, Carbon-based nanostructures as a versatile platform for tunable π-magnetism. *J. Phys.: Condens. Matter* **34**, 443001 (2022).

7. J. Brede, N. Merino-Díez, A. Berdonces-Layunta, S. Sanz, A. Domínguez-Celorrio, J. Lobo-Checa, M. Vilas-Varela, D. Peña, T. Frederiksen, J. I. Pascual, D. G. De Oteyza, D. Serrate, Detecting the spin-polarization of edge states in graphene nanoribbons. *Nat Commun* **14**, 6677 (2023).

8. S. Mishra, D. Beyer, K. Eimre, S. Kezilebieke, R. Berger, O. Gröning, C. A. Pignedoli, K. Müllen, P. Liljeroth, P. Ruffieux, X. Feng, R. Fasel, Topological frustration induces unconventional magnetism in a nanographene. *Nat. Nanotechnol.* **15**, 22–28 (2020).

9. C. Zhao, L. Yang, J. C. G. Henriques, M. Ferri-Cortés, G. Catarina, C. A. Pignedoli, J. Ma, X. Feng, P. Ruffieux, J. Fernández-Rossier, R. Fasel, Spin excitations in nanographene-based antiferromagnetic spin-1/2 Heisenberg chains. *Nat. Mater.*, doi: 10.1038/s41563-025-02166-1 (2025).

10. U. Ham, W. Ho, Imaging single electron spin in a molecule trapped within a nanocavity of tunable dimension. *The Journal of Chemical Physics* **138**, 074703 (2013).

11. L. L. Patera, S. Sokolov, J. Z. Low, L. M. Campos, L. Venkataraman, J. Repp, Resolving the Unpaired-Electron Orbital Distribution in a Stable Organic Radical by Kondo Resonance Mapping. *Angew Chem Int Ed* **58**, 11063–11067 (2019).

12. S. Baumann, W. Paul, T. Choi, C. P. Lutz, A. Ardavan, A. J. Heinrich, Electron paramagnetic resonance of individual atoms on a surface. *Science* **350**, 417–420 (2015).

13. K. Yang, W. Paul, S.-H. Phark, P. Willke, Y. Bae, T. Choi, T. Esat, A. Ardavan, A. J. Heinrich, C. P. Lutz, Coherent spin manipulation of individual atoms on a surface. *Science* **366**, 509–512 (2019).

14. Y. Chen, Y. Bae, A. J. Heinrich, Harnessing the Quantum Behavior of Spins on Surfaces. *Advanced Materials* **35**, 2107534 (2023).

15. T. S. Seifert, S. Kovarik, D. M. Juraschek, N. A. Spaldin, P. Gambardella, S. Stepanow, Longitudinal and transverse electron paramagnetic resonance in a scanning tunneling microscope. *Sci. Adv.* **6**, eabc5511 (2020).

16. X. Zhang, C. Wolf, Y. Wang, H. Aubin, T. Bilgeri, P. Willke, A. J. Heinrich, T. Choi, Electron spin resonance of single iron phthalocyanine molecules and role of their non-localized spins in magnetic interactions. *Nat. Chem.* **14**, 59–65 (2022).

17. S. Kovarik, R. Robles, R. Schlitz, T. S. Seifert, N. Lorente, P. Gambardella, S. Stepanow, Electron Paramagnetic Resonance of Alkali Metal Atoms and Dimers on Ultrathin MgO. *Nano Lett.* **22**, 4176–4181 (2022).

18. S. Reale, J. Hwang, J. Oh, H. Brune, A. J. Heinrich, F. Donati, Y. Bae, Electrically driven spin resonance of 4f electrons in a single atom on a surface. *Nat Commun* **15**, 5289 (2024).

19. G. Czap, K. Noh, J. Velasco, R. M. Macfarlane, H. Brune, C. P. Lutz, Direct Electrical Access to the Spin Manifolds of Individual Lanthanide Atoms. *ACS Nano*, acsnano.4c14327 (2025).

20. S. N. Datta, A. K. Pal, A. Panda, Design of magnetic organic molecules and organic magnets: Experiment, theory and computation with application and recent advances. *Chemical Physics Impact* **7**, 100379 (2023).

21. R. Kawaguchi, K. Hashimoto, T. Kakudate, K. Katoh, M. Yamashita, T. Komeda, Spatially Resolving Electron Spin Resonance of PI-Radical in Single-Molecule Magnet. *Nano Letters* **23**, 213–219 (2023).

22. T. Esat, D. Borodin, J. Oh, A. J. Heinrich, F. S. Tautz, Y. Bae, R. Temirov, A quantum sensor for atomic-scale electric and magnetic fields. *Nat. Nanotechnol.* **19**, 1466–1471 (2024).

23. S. Kovarik, R. Schlitz, A. Vishwakarma, D. Ruckert, P. Gambardella, S. Stepanow, Spin torque–driven electron paramagnetic resonance of a single spin in a pentacene molecule. *Science* **384**, 1368–1373 (2024).

24. L. Sellies, R. Spachtholz, S. Bleher, J. Eckrich, P. Scheuerer, J. Repp, Single-molecule electron spin resonance by means of atomic force microscopy. *Nature* **624**, 64–68 (2023).

25. W. Paul, K. Yang, S. Baumann, N. Romming, T. Choi, C. P. Lutz, A. J. Heinrich, Control of the millisecond spin lifetime of an electrically probed atom. *Nature Phys* **13**, 403–407 (2017).

26. A. Atto, A. Hudson, R. A. Jackson, N. P. C. Simmons, ESR of Fluorenyl and Identyl: Two Neutral Non-Alternant Radicals. *Chem. Phys. Lett.* **33**, 477 (1975).

27. D. R. Dalton, S. A. Liebman, Electron spin resonance studies on neutral aromatic hydrocarbon radicals. *Journal of the American Chemical Society* (1969).

28. Y. Tian, K. Uchida, H. Kurata, Y. Hirao, T. Nishiuchi, T. Kubo, Design and Synthesis of New Stable Fluorenyl-Based Radicals. *J. Am. Chem. Soc.* **136**, 12784–12793 (2014).

29. S. Mishra, S. Fatayer, S. Fernández, K. Kaiser, D. Peña, L. Gross, Nonbenzenoid High-Spin Polycyclic Hydrocarbons Generated by Atom Manipulation. *ACS Nano* **16**, 3264–3271 (2022).

30. S. Song, N. Guo, X. Li, G. Li, Y. Haketa, M. Telychko, J. Su, P. Lyu, Z. Qiu, H. Fang, X. Peng, J. Li, X. Wu, Y. Li, C. Su, M. J. Koh, J. Wu, H. Maeda, C. Zhang, J. Lu, Real-Space Imaging of a Single-Molecule Monoradical Reaction. *J. Am. Chem. Soc.* **142**, 13550–13557 (2020).

31. R. S. Klausen, J. R. Widawsky, T. A. Su, H. Li, Q. Chen, M. L. Steigerwald, L. Venkataraman, C. Nuckolls, Evaluating atomic components in fluorene wires. *Chem. Sci.* **5**, 1561 (2014).

32. G. Pacchioni, H. Freund, Electron Transfer at Oxide Surfaces. The MgO Paradigm: from Defects to Ultrathin Films. *Chem. Rev.* **113**, 4035–4072 (2013).

33. P. Willke, K. Yang, Y. Bae, A. J. Heinrich, C. P. Lutz, Magnetic resonance imaging of single atoms on a surface. *Nat. Phys.* **15**, 1005–1010 (2019).

34. H. K. Fun, K. Sivakumar, D. R. Zhu, X. Z. You, 4,5-Diazafluoren-9-one. *Acta Crystallogr C Cryst Struct Commun* **51**, 2076–2078 (1995).

35. C. Chiang, C. Xu, Z. Han, W. Ho, Real-space imaging of molecular structure and chemical bonding by single-molecule inelastic tunneling probe. *Science* **344**, 885–888 (2014).

36. Z. Han, G. Czap, C. Chiang, C. Xu, P. J. Wagner, X. Wei, Y. Zhang, R. Wu, W. Ho, Imaging the halogen bond in self-assembled halogenbenzenes on silver. *Science* **358**, 206–210 (2017).

37. L. Gross, F. Mohn, N. Moll, P. Liljeroth, G. Meyer, The Chemical Structure of a Molecule Resolved by Atomic Force Microscopy. *Science* **325**, 1110–1114 (2009).

38. P. Hapala, G. Kichin, C. Wagner, F. S. Tautz, R. Temirov, P. Jelínek, Mechanism of high-resolution STM/AFM imaging with functionalized tips. *Phys. Rev. B* **90**, 085421 (2014).

39. P. Hapala, R. Temirov, F. S. Tautz, P. Jelínek, Origin of High-Resolution IETS-STM Images of Organic Molecules with Functionalized Tips. *Phys. Rev. Lett.* **113**, 226101 (2014).

40. B. De La Torre, M. Švec, G. Foti, O. Krejčí, P. Hapala, A. Garcia-Lekue, T. Frederiksen, R. Zbořil, A. Arnau, H. Vázquez, P. Jelínek, Submolecular Resolution by Variation of the Inelastic Electron Tunneling Spectroscopy Amplitude and its Relation to the AFM/STM Signal. *Phys. Rev. Lett.* **119**, 166001 (2017).

41. F. E. Olsson, S. Paavilainen, M. Persson, J. Repp, G. Meyer, Multiple Charge States of Ag Atoms on Ultrathin NaCl Films. *Phys. Rev. Lett.* **98**, 176803 (2007).

42. K. Yang, W. Paul, F. D. Natterer, J. L. Lado, Y. Bae, P. Willke, T. Choi, A. Ferrón, J. Fernández-Rossier, A. J. Heinrich, C. P. Lutz, Tuning the Exchange Bias on a Single Atom from 1 mT to 10 T. *Phys. Rev. Lett.* **122**, 227203 (2019).

43. K. Yang, Y. Bae, W. Paul, F. D. Natterer, P. Willke, J. L. Lado, A. Ferrón, T. Choi, J. Fernández-Rossier, A. J. Heinrich, C. P. Lutz, Engineering the Eigenstates of Coupled Spin- 1 / 2 Atoms on a Surface. *Phys. Rev. Lett.* **119**, 227206 (2017).

44. D. Rugar, R. Budakian, H. J. Mamin, B. W. Chui, Single spin detection by magnetic resonance force microscopy. *Nature* **430**, 329–332 (2004).

45. B. Verlhac, N. Bachellier, L. Garnier, M. Ormaza, P. Abufager, R. Robles, M.-L. Bocquet, M. Ternes, N. Lorente, L. Limot, Atomic-scale spin sensing with a single molecule at the apex of a scanning tunneling microscope. *Science* **366**, 623–627 (2019).

46. G. Czap, P. J. Wagner, F. Xue, L. Gu, J. Li, J. Yao, R. Wu, W. Ho, Probing and imaging spin interactions with a magnetic single-molecule sensor. *Science* **364**, 670–673 (2019).

47. B. Schuler, W. Liu, A. Tkatchenko, N. Moll, G. Meyer, A. Mistry, D. Fox, L. Gross, Adsorption Geometry Determination of Single Molecules by Atomic Force Microscopy. *Phys. Rev. Lett.* **111**, 106103 (2013).

48. M. Ternes, Spin excitations and correlations in scanning tunneling spectroscopy. *New J. Phys.* **17**, 063016 (2015).

49. S. Loth, C. P. Lutz, A. J. Heinrich, Spin-polarized spin excitation spectroscopy. *New J. Phys.* **12**, 125021 (2010).

50. J. Reina-Gálvez, C. Wolf, N. Lorente, Many-body nonequilibrium effects in all-electric electron spin resonance. *Phys. Rev. B* **107**, 235404 (2023).

51. P. Kot, M. Ismail, R. Drost, J. Siebrecht, H. Huang, C. R. Ast, Electric control of spin transitions at the atomic scale. *Nat Commun* **14**, 6612 (2023).

52. X. Zhang, J. Reina-Gálvez, D. Wu, J. Martinek, A. J. Heinrich, T. Choi, C. Wolf, Electric field control of the exchange field of a single spin impurity on a surface.

53. G. Mikaelian, N. Ogawa, X. W. Tu, W. Ho, Atomic scale control of single molecule charging. *The Journal of Chemical Physics* **124**, 131101 (2006).

54. G. V. Nazin, S. W. Wu, W. Ho, Tunneling rates in electron transport through double-barrier molecular junctions in a scanning tunneling microscope. *Proc. Natl. Acad. Sci. U.S.A.* **102**, 8832–8837 (2005).

55. N. Krane, C. Lotze, N. Bogdanoff, G. Reecht, L. Zhang, A. L. Briseno, K. J. Franke, Mapping the perturbation potential of metallic and dipolar tips in tunneling spectroscopy on MoS 2. *Phys. Rev. B* **100**, 035410 (2019).

56. M. Ellner, N. Pavliček, P. Pou, B. Schuler, N. Moll, G. Meyer, L. Gross, R. Peréz, The Electric Field of CO Tips and Its Relevance for Atomic Force Microscopy. *Nano Lett.* **16**, 1974–1980 (2016).

57. S. Duan, G. Tian, X. Xu, A General Framework of Scanning Tunneling Microscopy Based on Bardeen's Approximation for Isolated Molecules. *JACS Au* **3**, 86–92 (2023).

58. A. M. Shakirov, A. N. Rubtsov, P. Ribeiro, Spin transfer torque induced paramagnetic resonance. *Phys. Rev. B* **99**, 054434 (2019).

59. S. Loth, K. Von Bergmann, M. Ternes, A. F. Otte, C. P. Lutz, A. J. Heinrich, Controlling the state of quantum spins with electric currents. *Nature Phys* **6**, 340–344 (2010).

60. W. Paul, S. Baumann, C. P. Lutz, A. J. Heinrich, Generation of constant-amplitude radio-frequency sweeps at a tunnel junction for spin resonance STM. *Review of Scientific Instruments* **87**, 074703 (2016).

61. M. Steinbrecher, W. M. J. Van Weerdenburg, E. F. Walraven, N. P. E. Van Mullekom, J. W. Gerritsen, F. D. Natterer, D. I. Badrtdinov, A. N. Rudenko, V. V. Mazurenko, M. I. Katsnelson, A. Van Der Avoird, G. C. Groenenboom, A. A. Khajetoorians, Quantifying the interplay between fine structure and geometry of an individual molecule on a surface. *Phys. Rev. B* **103**, 155405 (2021).

62. M. W. Schmidt, K. K. Baldridge, J. A. Boatz, S. T. Elbert, M. S. Gordon, J. H. Jensen, S. Koseki, N. Matsunaga, K. A. Nguyen, S. Su, T. L. Windus, M. Dupuis, J. A. Montgomery, General atomic and molecular electronic structure system. *J Comput Chem* **14**, 1347–1363 (1993).

63. P. Giannozzi, S. Baroni, N. Bonini, M. Calandra, R. Car, C. Cavazzoni, D. Ceresoli, G. L. Chiarotti, M. Cococcioni, I. Dabo, A. Dal Corso, S. De Gironcoli, S. Fabris, G. Fratesi, R. Gebauer, U. Gerstmann, C. Gougoussis, A. Kokalj, M. Lazzeri, L. Martin-Samos, N. Marzari, F. Mauri, R. Mazzarello, S. Paolini, A. Pasquarello, L. Paulatto, C. Sbraccia, S. Scandolo, G. Sclauzero, A. P. Seitsonen, A. Smogunov, P. Umari, R. M. Wentzcovitch, QUANTUM ESPRESSO: a modular and open-source software project for quantum simulations of materials. *J. Phys.: Condens. Matter* **21**, 395502 (2009).

64. P. Giannozzi, O. Andreussi, T. Brumme, O. Bunau, M. Buongiorno Nardelli, M. Calandra, R. Car, C. Cavazzoni, D. Ceresoli, M. Cococcioni, N. Colonna, I. Carnimeo, A. Dal Corso, S. De Gironcoli, P. Delugas, R. A. DiStasio, A. Ferretti, A. Floris, G. Fratesi, G. Fugallo, R. Gebauer, U. Gerstmann, F. Giustino, T. Gorni, J. Jia, M. Kawamura, H.-Y. Ko, A. Kokalj, E. Küçükbenli, M. Lazzeri, M. Marsili, N. Marzari, F. Mauri, N. L. Nguyen, H.-V. Nguyen, A. Otero-de-la-Roza, L. Paulatto, S. Poncé, D. Rocca, R. Sabatini, B. Santra, M. Schlipf, A. P. Seitsonen, A. Smogunov, I. Timrov, T. Thonhauser, P. Umari, N. Vast, X. Wu, S. Baroni, Advanced capabilities for materials modelling with Quantum ESPRESSO. *J. Phys.: Condens. Matter* **29**, 465901 (2017).

65. A. Dal Corso, Pseudopotentials periodic table: From H to Pu. *Computational Materials Science* **95**, 337–350 (2014).

66. R. Sabatini, T. Gorni, S. De Gironcoli, Nonlocal van der Waals density functional made simple and efficient. *Phys. Rev. B* **87**, 041108 (2013).

67. M. Braun, J. König, J. Martinek, Theory of transport through quantum-dot spin valves in the weak-coupling regime. *Phys. Rev. B* **70**, 195345 (2004).

68. J. Reina-Galvez, M. Nachtigall, N. Lorente, J. Martinek, C. Wolf, Contrasting exchange-field and spin-transfer torque driving mechanisms in all-electric electron spin resonance. arXiv arXiv:2503.24046 [Preprint] (2025). https://doi.org/10.48550/arXiv.2503.24046.

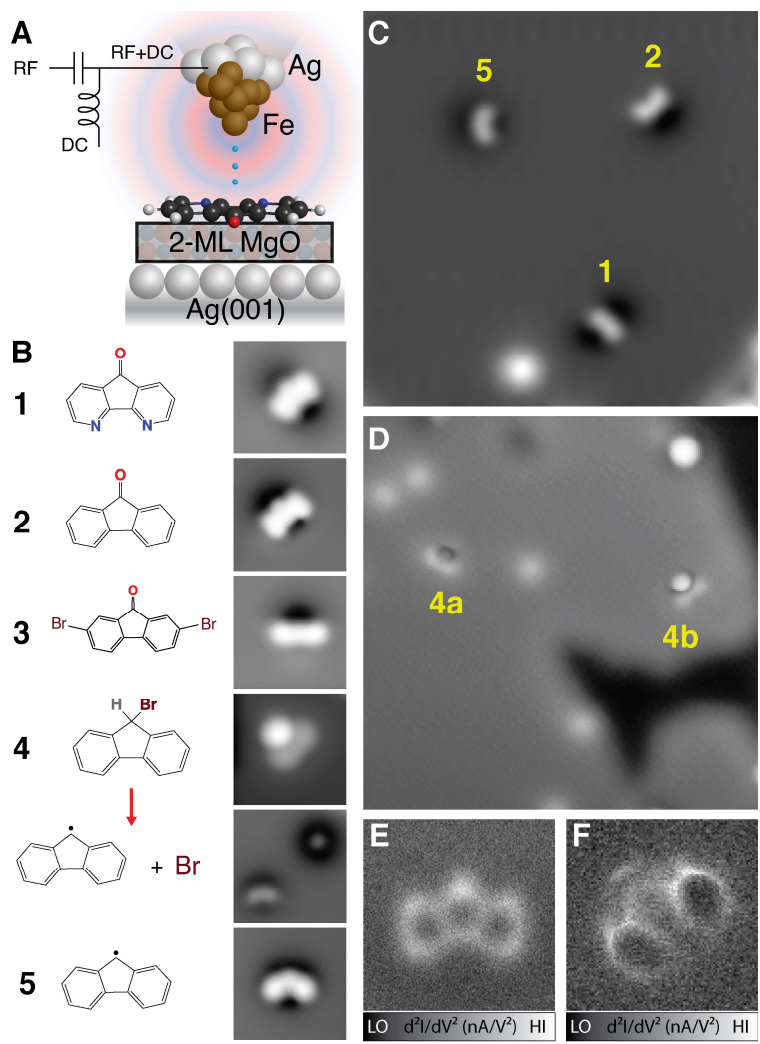

**Figure 1: Organic molecules probed by STM.** (**A**) Schematic diagram depicting the ESR-STM experiment, where molecules are adsorbed on a thin epitaxial MgO film. (**B**) Chemical structure diagrams and corresponding STM topographs of the molecular species studied in this work: (**1**) 4,5-diazafluorenone (DAF), (**2**) 9-fluorenone, (**3**) 2,7-dibromo-9-fluorenone (DBF), (**4**) 9-bromo-fluorene with dissociation products shown, and (**5**) 9-fluorenyl radical. Each topograph is 2−3.4 nm square, acquired with Br-terminated tips to enhance spatial resolution. (**C**) STM constant-current image of co-adsorbed DAF, fluorenone, and fluorenyl on 2-ML MgO/Ag(001). Imaging conditions 20 pA at −60 mV, 10 nm square, metal-terminated tip. (**D**) STM topograph of bromofluorene molecules adsorbed with the Br atom oriented toward the surface (4a) or away from the surface (4b). Imaging conditions 20 pA at +150 mV, 10 nm square, Br-terminated tip. (**E,F**) Bond-resolved images of fluorenone (E) and DAF (F) adsorbed directly on Ag(001), acquired with a CO-terminated tip using the inelastic tunneling probe (itProbe) method (*35, 36*). Images show $d^2I/dV^2$ maps at constant height, tip-height setpoint 400 pA at 20 mV on molecule center, bias modulation $V_{mod}$ = 1 mV r.m.s. at 803 Hz, 1.2 nm square. $V_{DC}$ = 2 mV for (E), 1.5 mV for (F).

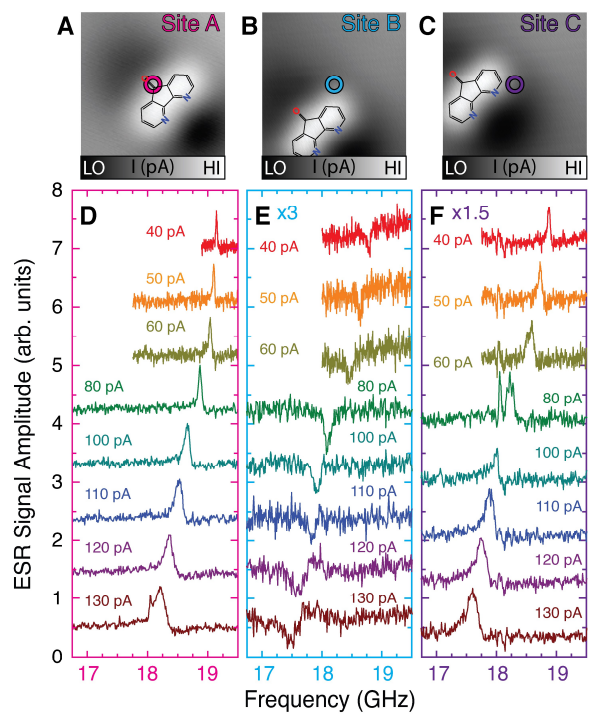

**Figure 2: ESR spectra of diazafluorenone (DAF)**. (**A**–**C**) Tunnel-current STM images of DAF acquired at constant tip height. Images are centered on three representative sites (colored circles). Tip height was established with tip positioned over molecule center with setpoint $I_{set}$ = 80 pA at $V_{set}$ = 50 mV before opening the constant-current feedback loop. (**D**–**F**) ESR spectra acquired with tip positioned at the respective sites shown in (A) to (C). All spectra: tip-height setpoint $V_{set}$ = 60 mV and current $I_{set}$ as labeled with tip above molecule center; $V_{DC}$ = +100 mV during spectrum acquisition. $V_{RF}$ = 30 mV 0-p (zero-to-peak), $B$ = 0.7 T.

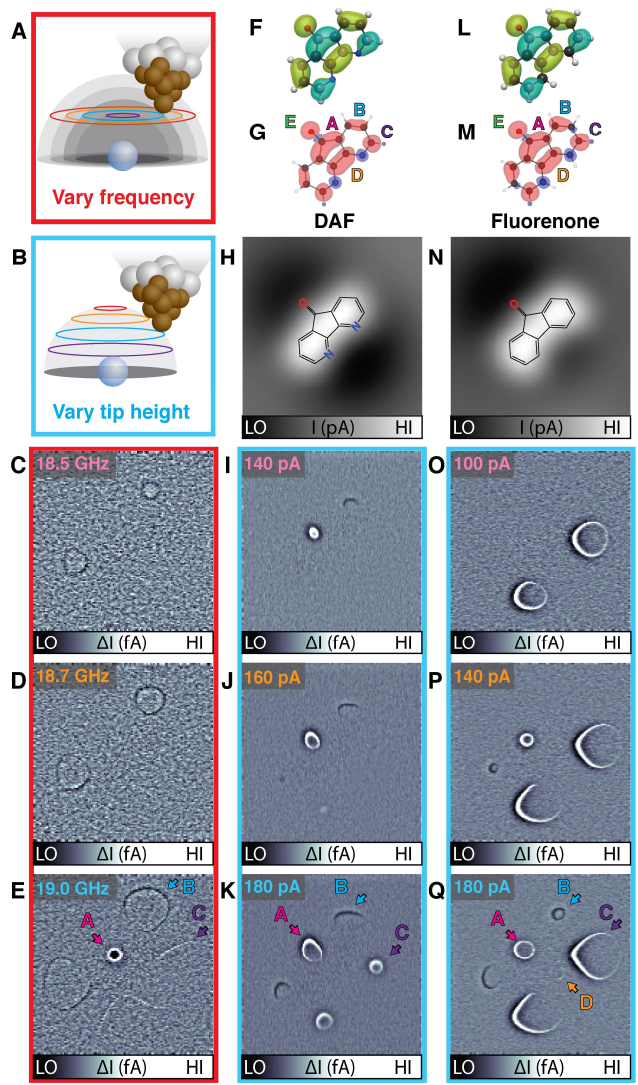

**Figure 3: Scanning magnetic resonance images (MRI) of DAF and fluorenone.** (**A**) Diagram of Scheme 1 for MRI sequences, in which the tip-surface distance is held constant and the RF frequency $f$ is varied between successive MRI images. This results in images that probe different tip-molecule interaction energy isosurfaces (concentric grey spheres) in a 2D imaging plane (colored rings). (**B**) Scheme 2 MRI, in which the tip-surface distance is varied between successive MRI images. This maps a single 3D tip-adsorbate exchange energy isosurface as a series of 2D slices. (**C**–**E**) Scheme 1 MRI sequence of a DAF molecule at three different RF frequencies as labeled. The Zeeman energy at $B = 0.7$ T applied here is ~19.5 GHz. The constant-current feedback loop was open during MRI acquisition, using an initial tip-height setpoint 135 pA at 40 mV with tip positioned at molecule center. (**F**, **G**) DFT-calculated properties of gas-phase DAF anion. (F) SOMO wavefunction isosurfaces; blue and green regions having opposite wavefunction sign. (G) Spin density isosurfaces. Pink: majority spin; blue: minority spin. Colored letters label the distinct spin-density lobes. (**H**) Tunnel current image of DAF acquired concurrently with the MRI image of (K). (**I**–**K**) Scheme 2 MRI sequence of DAF at three different tip-surface distances that are ~60 pm apart, established by setting tunnel current at molecule center as labeled at $V_{set}$ = 60 mV. Imaging conditions: $V_{DC}$ = 100 mV, $f$ = 24 GHz. (**L, M**) DFT-calculated SOMO wavefunction (L) and spin density (M) of gas-phase fluorenone anion. (**N–Q**) Tunnel current image (N) and Scheme 2 MRI images (~120 pm apart in $z$) of a fluorenone molecule (O–Q) acquired using the same tip apex and imaging conditions as (H–K). Images are rotated 90° to aid comparison with (H–K). All MRI images: 1.5 nm square, $V_{RF}$ = 30 mV, $B$ = 0.7 T. All MRI images are high-pass filtered for clarity (see SM).

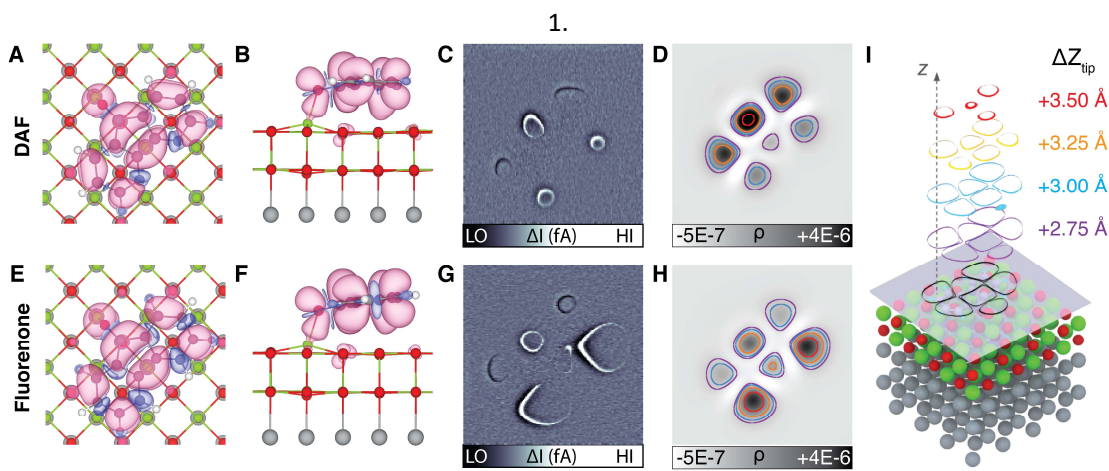

**Figure 4: DFT calculations of adsorbed DAF and fluorenone.** (**A**) Top and (**B**) side views of the relaxed geometry of DAF in the diagonal orientation bound to 2 ML MgO/Ag(001). Pink (blue) regions show majority (minority) spin-density isosurfaces, with O atoms (green balls), Mg (red) and Ag (grey). (**C**) Constant-height MRI image of DAF. Image is 1.5 nm square, tip height setpoint 200 pA at 60 mV, imaging conditions $V_{DC}$ = 100 mV, $B$ = 0.7 T. (**D**) DFT-calculated plot of spin density ρ of DAF in a plane parallel to surface. Colored lines are contours of constant spin density. Image 1.5 nm square. (**E**) Top and (**F**) side views of fluorenone in its relaxed adsorption geometry. (**G**) Constant-height MRI image of fluorenone; conditions as in (C). (**H**) DFT-calculated spin-density plot of fluorenone; conditions as in (D). (**I**) Perspective plot of spin density contours of DAF at tip heights as labeled. Heights are distances to the plane of Mg atom centers of the top MgO plane. Colors correspond to contour colors in (D).

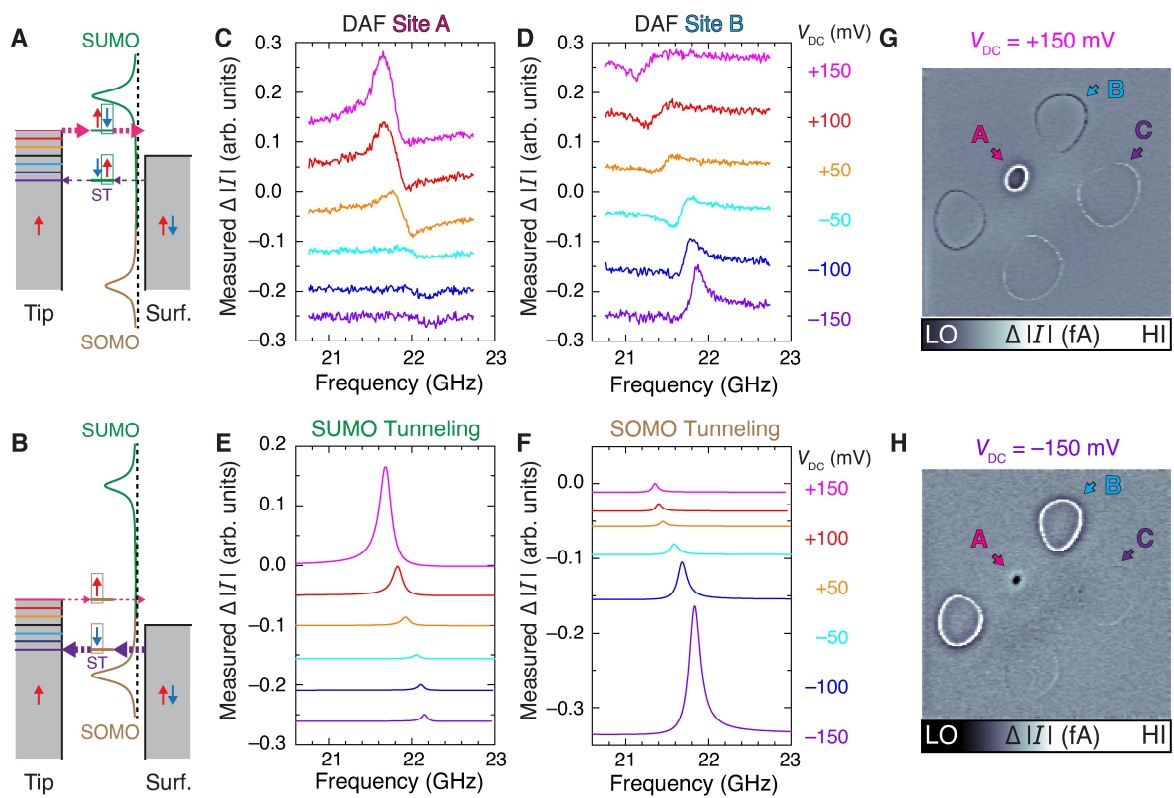

**Figure 5: Effects of bias voltage and orbital properties on ESR image contrast.** (**A**, **B**) Energy level diagrams showing SOMO and SUMO orbital resonances (curves) and tunneling processes to illustrate mechanisms that favor either SUMO tunneling (A) or SOMO tunneling (B). In (A), electrostatic gating by the STM tip shifts both levels down in energy to favor SUMO tunneling. Independently, SUMO local density of states is higher in (A) than in (B) to favor SUMO tunneling. More positive sample bias voltage (red) favors SUMO tunneling in both (A) and (B). Horizontal arrows show tunneling processes for positive (red) and negative (purple) bias. For negative bias, "ST" indicates that the initial (off-resonance) population is inverted due to spin torque. Vertical arrows indicate spin orientations in each orbital; boxed arrow indicates the state and spin orientation for the dominant tunneling channel. (**C**, **D**) ESR spectra of DAF for bias voltages $V_{DC}$ indicated. Tip is positioned where tunneling is dominated by SUMO (C) or SOMO (D) tunneling. Signals shown are changes in absolute magnitude of current. Setpoint $I_{set}$ = 160 pA at $V_{set}$ = +100 mV, $B$ = 0.85 T. (**E**, **F**) Simulated ESR signals corresponding to SUMO and SOMO tunneling using non-equilibrium transport model described in main. (**G**, **H**) MRI images of DAF with $V_{DC}$ = +100 mV (G) and $V_{DC}$ = −100 mV (H). Lighter color represents higher magnetoconductance (greater current magnitude) for both bias polarities. Reversal of conductance contrast with sign of $V_{DC}$ is due to spin torque initialization, which inverts the off–ESR-resonance molecular spin when $V_{DC}$ < 0. Tip-height setpoint is 240 pA at 100 mV at molecule center. Images 1.5 nm square, $f$ = 21.5 GHz, $B$ = 0.85 T.